\documentclass[twoside,fleqn]{ActaStyle}
\usepackage{times}

\usepackage{lscape}
\usepackage{graphicx,epsfig,psfrag}
\usepackage[tbtags]{amsmath}

\newcommand{\ba}{\begin{eqnarray}}
\newcommand{\ea}{\end{eqnarray}}
\newcommand{\bea}{\begin{array}{l}}
\newcommand{\eea}{\end{array}}
\newcommand{\ldl}{\Lambda \partial_{\Lambda}}
\newcommand{\ie}{{\it i.e.}\ }
\newcommand{\eg}{{\it e.g.}\ }

\newcommand{\cf}{{\it cf.}\ }
\newcommand{\tr}{\mathrm{tr}}
\newcommand{\Lam}{\Lambda}
\newcommand{\C}{{\cal C}}
\newcommand{\A}{{\cal A}}

\def\sqr#1#2{{\vcenter{\vbox{\hrule height .#2pt
        \hbox{\vrule width .#2pt height#1pt \kern#1pt
              \vrule width.#2pt}
          \hrule height .#2pt}}}}

\def\hepth#1{{\tt hep-th/#1}}

\newcommand{\bSUNN}{$\boldsymbol{SU(N|N)}$\ }
\newcommand{\bSUN}{$\boldsymbol{SU(N)}$\ }
\newcommand{\str}{{\mathrm{str}}}
\newcommand{\one}{\! \hbox{ 1\kern-.8mm l}}
\newcommand{\ct}{{\tilde c}}
\def\dphi#1#2{\frac{\delta #1}{\delta \varphi_{#2}} }
\newcommand{\demu}{\partial_{\mu}}

\def\authorheading{S.~Arnone et al.}

\def\shorttitle{Gauge invariant ERG}

\begin{document}
%% ---------------------------------------------------

%%%  page range, first and last page
\pagerange{1}{8}

%%% paper title
\title{%
A MANIFESTLY GAUGE INVARIANT EXACT  \\
RENORMALIZATION GROUP
}

%%% author(s) and address(es)
\author{%  author(s)
Stefano Arnone\email{S.Arnone@soton.ac.uk}, 
Antonio Gatti\email{A.Gatti@soton.ac.uk}, 
Tim R. Morris\email{T.R.Morris@soton.ac.uk}
}
{%  address(es)
Department of Physics and Astronomy, University of Southampton\\ 
Highfield, Southampton SO17 1BJ, United Kingdom.
}

%%% Date of submition
\day{May 14, 2002}

%%% abstract of the paper
\abstract{%
A manifestly gauge invariant exact renormalization group for pure
$SU(N)$ Yang-Mills theory is proposed, allowing gauge invariant
calculations, without any gauge fixing or ghosts. The necessary gauge
invariant regularisation which implements the effective cutoff, is
naturally incorporated %A novel regularisation scheme, implemented 
by embedding the theory into a spontaneously broken $SU(N|N)$ super-gauge
theory. This guarantees finiteness to all orders in perturbation
theory.
}

%%% PASC numbers of your article

% %%%%%%%%%%%%%%%%%%%%%%%%%%%%%%%%%%%%%%%%%%%%%%%%%%%%%%%%%%%%%%%%%
\section{ERG and gauge invariance}
% %%%%%%%%%%%%%%%%%%%%%%%%%%%%%%%%%%%%%%%%%%%%%%%%%%%%%%%%%%%%%%%%%
\label{sec:erg} \setcounter{section}{1}\setcounter{equation}{0}

The basic idea of the exact renormalization group (ERG) is summarised in the
diagram below. For a detailed review, and current developments,
see for example \cite{YKIS,devel}. In the
partition function for the theory, defined in the continuum and in
Euclidean space, rather than integrate over all momentum modes in one
go, one first integrates out modes between an overall cutoff
$\Lambda_0$ and the effective Wilsonian cutoff $\Lambda << \Lambda_0$.
The remaining integral can again be expressed as a partition function,
but the bare action, $S_{\Lambda_0}$, is replaced by an effective
action, $S$. The new Boltzmann
factor, $\exp -S$, is more or less the original partition function,
modified by an infrared cutoff $\Lambda$ \cite{ergi}. When finally $\Lambda$ is
sent to zero, the full partition function is recovered and all the
physics that goes with it (\eg Green functions). In practice however,
one does not work with this integral form but rather a differential 
equation for $S$, the ERG equation, that expresses how $S$
changes as one lowers $\Lambda$.

%  Fig. 1
\begin{figure}[t]
\begin{center}
\includegraphics[width=6.5cm]{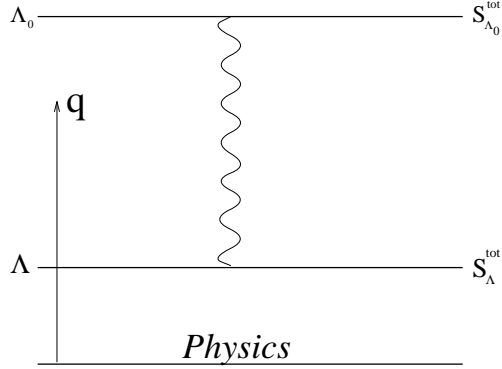} \\
\end{center}
\caption{%
Caption of figure.
}
\label{fig:1}
\end{figure}

The application of this 
technique to quantum field theory brings with it many advantages, because 
renormalization properties,
which are normally subtle and complicated, are here 
--using Wilson's insight \cite{kogwil}--
straightforward to build in from the beginning. Thus solutions
for the effective action may be found directly in terms of renormalized
quantities (in fact without specifying  a bare action at $\Lambda_0$,
which is anyway, by universality, largely arbitrary), and within
this framework almost any
approximation can be considered (for example
truncations \cite{trunc}, derivative expansion \cite{deriv} etc.) 
without disturbing this property \cite{YKIS}. 
The result is that these ideas form a powerful framework for considering 
non-perturbative analytic approximations in quantum field theory \cite{devel}.

In particle physics, all the interesting non-perturbative questions also
involve gauge theory.  However, in order to 
construct a gauge invariant ERG,
we must overcome an obvious conflict: the
division of momenta into large and small, according to the effective
scale $\Lambda$, is not preserved by gauge transformations. (Explicitly,
consider a matter field $\phi(x)$. Under a gauge transformation
$\phi(x) \rightarrow \Omega(x) \, \phi(x)$ momentum modes $\phi(p)$ are
mapped to a convolution with the modes from $\Omega$.) 
We only have two choices.
Either we break the gauge invariance and try to recover it once
the cutoff is removed, by imposing suitable boundary conditions on the ERG 
equation \cite{pawl}, or we generalise things so that we can
write down a gauge invariant ERG equation.
 
We will go with the second choice \cite{alg,ymi,ymii}. 
Furthermore, we will find
that we can continue to keep the gauge invariance manifest at all
stages even when we start to compute the effective action. No gauge
fixing or ghosts are required. The full power and beauty of gauge
invariance thus shines through in the calculations, as will all become
clear in the next talk \cite{sa}.

% %%%%%%%%%%%%%%%%%%%%%%%%%%%%%%%%%%%%%%%%%%%%%%%%%%%%%%%%%%%%%%%%%
\section{Regularisation via \bSUNN}
% %%%%%%%%%%%%%%%%%%%%%%%%%%%%%%%%%%%%%%%%%%%%%%%%%%%%%%%%%%%%%%%%%
\label{sec:reg}
\subsection{General idea}
\label{subsec:gen}

As a necessary first step, we need a gauge invariant  implementation 
of the non-perturbative continuum effective cutoff $\Lambda$. 
The standard ERG cutoff is implemented by inserting 
$c^{-1}(p^2/\Lambda^2)$ into the kinetic term of the action.
$c$ is a smooth ultraviolet cutoff profile with $c(0)=1$, 
decaying sufficiently rapidly as $p/\Lambda\to\infty$ that all 
quantum corrections are regularised. To restore the gauge invariance we
covariantise so that the regularised bare action takes the form:
\be
\label{eq:1}
\frac{1}{2g^2}\, \tr\! \int\! {\rm d}^{D}\!x \; F_{\mu \nu}  \, c^{-1} 
\left( -{D^2}/{\Lambda^2} \right)  \cdot F^{\mu \nu}.
\ee
Here $F_{\mu\nu}=i[D_\mu,D_\nu]$ is the standard field strength,
built from the covariant derivative $D_\mu=\partial_\mu-i A_\mu$.
We scale out the coupling $g$ for good reason: since gauge invariance
will be exactly preserved, the form of the covariant derivative is
protected \cite{rome}, which in this parametrisation simply means  that $A$
suffers no wavefunction renormalization. Eq.~(\ref{eq:1}) is nothing but
covariant higher derivative regularisation and is known to fail at
one-loop \cite{olfail}. Slavnov solved this problem by introducing
gauge invariant Pauli-Villars fields \cite{slav}. These appear bilinearly 
so that their one-loop determinants cancel the remaining divergences.
We cannot use these ideas directly since the bilinearity
property cannot be preserved by
the ERG flow \cite{alg,ymii}. 
Instead, we discovered a novel and elegant solution:
we embed (\ref{eq:1}) in a spontaneously
broken $SU(N|N)$ super-gauge theory \cite{sunn}. We will see that
the result has similar characteristics to Slavnov's scheme but
sits much more naturally in the effective action framework. Indeed
the regularising properties will follow from the supersymmetry in
the fibres of the high energy unbroken supergroup. We will then 
design an ERG in which the spontaneous
breaking scale and higher derivative scale are identified and
flow together as we lower $\Lambda$.

\subsection{The \bSUNN super group}
\label{subsec:sunn}

The graded Lie algebra of $SU(N|M)$ in the $(N+M)$ - dimensional
representation  is given by 
\be
{\cal H} = \left( \begin{array}{cc} H_N & \theta \\
                               \theta^\dagger & H_M	
               \end{array}
       \right).
\ee		
$H_N \, (H_M)$ is an $N\times N \, (M \times M)$ Hermitian matrix with 
complex bosonic elements and  $\theta$ is an
$M\times N$ matrix composed of  complex Grassmann numbers.
${\cal H}$ is required to be supertraceless, \ie
\be
\label{eq:str}
\str({\cal H})= \tr(\sigma_3{\cal H})
= \tr(H_N) - \tr(H_M)
=0  
\ee
(where $\sigma_3={\rm diag}(\one_N,-\one_M)$ is the obvious generalisation
of the Pauli matrix to this context).
The traceless parts of $H_N$ and $H_M$ correspond to $SU(N)$ and  $SU(M)$ 
respectively and the traceful part gives rise to a $U(1)$, so we see that 
the bosonic sector of the $SU(N|M)$ algebra  forms a  
$SU(N)\times SU(M) \times U(1)$ sub-algebra.

Specialising to $M=N$, we see that the $U(1)$ generator becomes just
$\one_{2N}$ and thus commutes with all the other generators. We cannot
simply drop it however because it is generated by other elements of
the algebra (\eg $\{\sigma_1,\sigma_1\}=2\one_{2N}$). Bars suggested 
removing it by redefining the Lie bracket to project out traceful
parts \cite{bars}:
$
[\ ,\ ]_{\pm}\mapsto[\ ,\ ]_{\pm}-{\one\over2N}\tr[\ ,\ ]_{\pm}.
$
We can use this idea but only on the gauge fields:
the matter fields require the full commutator because invariance
of the Lagrangian in this sector
requires the bracket to be Leibnitz \cite{sunn}. A simpler
and equivalent solution \cite{sunn} is to
keep the $\one_{2N}$ and note that the corresponding gauge field,
$\A^0$, which we have seen is needed to absorb gauge transformations 
produced in the $\one_{2N}$ direction,
does not however appear in the Lagrangian at all! (Its absence is then
protected by a no-$\A^0$ shift-symmetry: $\delta\A^0_\mu=\Lambda_\mu$.)

\subsection{Higher derivative \bSUNN super-gauge theory}
\label{subsec:high}

\begin{figure}[h]
\psfrag{dots}{$\cdots$}
\begin{center}
\includegraphics[scale=.4]{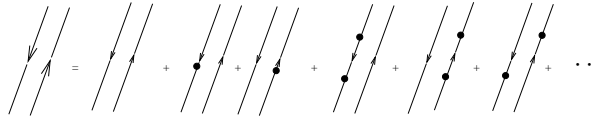} \\
\end{center}
\caption{%
Expansion of the covariantization in terms of super-gauge fields.
}
\label{fig:2}
\end{figure}

We promote the gauge field to a connection for $SU(N|N)$:
\be
\A_\mu = {\A}^{0}_{\mu} \one
+ \left( \!\! \begin{array}{cc}
                   A^{1}_{\mu} & B_{\mu} \\
                   \bar{B}_{\mu} & A^{2}_{\mu}
                   \end{array} \!\!
            \right),
\ee
where the $A^i_\mu$ are two the bosonic gauge fields for $SU(N)\times SU(N)$,
and $B_\mu$ is a fermionic gauge field. The field strength ${\cal F}_{\mu \nu}$
is now a commutator of the super-covariant derivative $\nabla_\mu=\partial_\mu-i\A_\mu$.
The super-gauge field part of the Lagrangian is then
\be
\label{eq:la}
{\cal L}_{\A} = {1\over 2g^2}{\cal F}_{\mu \nu} \{c^{-1}
\}{\cal F}^{\mu\nu}.
\ee
Here we take the opportunity to be more sophisticated about the 
covariantization of the cutoff. % $c$. 
For any momentum space kernel
$W(p^2/\Lambda^2)$, there are infinitely many covariantizations. The
form used in (\ref{eq:1}) is just one of them. Another way would be to
use Wilson lines \cite{rome,alg}.
In general, the covariantization results in a new set vertices
(infinite in number if $W$ is not a polynomial):
\ba
{\bf u} \{W\}  {\bf v} =
\sum_{n,m=0} &{}&\!\!\!\!\!\!\!\int_{x,y} \int_{x_i y_j} 
W_{\mu_1\cdots\mu_n,\nu_1\cdots\nu_m}(x_i;y_j ;x,y) \nonumber\\
&{}& \str [{\bf u}(x) {\cal A}_{\mu_1}(x_1)\cdots {\cal A}_{\mu_n}(x_n)
{\bf v}(y) {\cal A}_{\nu_1}(y_1)\cdots {\cal A}_{\nu_m}(y_m)]. 
\ea
(${\bf u}(x)$ and ${\bf v}(y)$ are
any two supermatrix representations.)
These can be graphically represented as in Fig.~\ref{fig:2}.

%  Fig. 2

\subsection{Spontaneous breaking in fermionic directions}
\label{subsec:spon}
Now we add a super-scalar field, ${\cal L} = {\cal L}_{\A} + {\cal L}_{\C}$,
\be
{\cal C} = \left( \!\! \begin{array}{cc}
                   C^{1} & D \\
                   \bar{D} & C^{2}
                   \end{array}  \!\!
            \right) \in U(N|N),
\ee
with a Lagrangian that encourages spontaneous symmetry breaking:
\be
{\cal L}_{\C} = {1\over 2} \nabla_{\mu}\cdot \C \{\ct^{-1}
\}\nabla^{\mu}\cdot \C  + 
{\lambda \over {4}} \str \! \int \! \Big(\C^2 - \Lam^{2}\Big)^2.
\ee
Choosing the classical vacuum expectation value $\Lambda\sigma_3$
breaks all and only the fermionic directions,
and expanding about this, by $\C\mapsto\C+\Lambda\sigma_3$, gives
\ba
{\cal L}_{\C} = &&\!\!\!
{1\over 2} \nabla_{\mu}\cdot \C \{\ct^{-1} \}\nabla^{\mu}\cdot \C
-i\Lambda [\A_\mu,\sigma_3]\{\ct^{-1}\}\nabla^{\mu}\cdot \C \nonumber\\
&&-{1\over 2}\Lam^2[\A_\mu,\sigma_3]\{\ct^{-1}\}[\A^{\mu},\sigma_3] 
+{\lambda \over 4} 
\str \!\! \int \!\! \Big(\Lambda \{\sigma_3,\C\} +\C^2\Big)^{\!2}. 
\ee
Since (fermionic) bosonic parts (anti)commute with $\sigma_3$, 
we see in the
second line that $B$ gains a mass $\sqrt{2}\Lambda$ ($B$ eats $D$),
and $C^1$ and $C^2$ gain masses $\sqrt{2\lambda}\Lambda$. 
These heavy fields play the r\^ole of Slavnov's
gauge invariant Pauli-Villars fields.

\subsection{Proof of regularisation}
\label{subsec:proof}

A proof that this all adds up to a regularisation of four
dimensional $SU(N)$ Yang-Mills
theory has been given in \cite{sunn}.
We only have room to summarise the conclusions. 

If $c^{-1}$ and $\ct^{-1}$
are chosen to be polynomials of rank $r$, $\tilde{r}$, we require
$r > \tilde{r}-1$ and $\tilde{r} > -1$ simply to ensure that at high 
momentum the propagators go over to those of the unbroken $SU(N|N)$ 
theory. The stronger constraints $\tilde{r}>1$ and $r-\tilde{r}>1$
then ensure finiteness in all perturbative diagrams except
pure $\A$ one-loop graphs with up to $4$ external legs. This maximises
the regularising power of the covariant higher derivatives and is ensured
simply by power counting.
The remaining diagrams can be shown to be finite 
within spontaneously broken $SU(N|N)$ gauge
theory as follows.
One-loop diagrams with 2 or 3 external $\A$ legs are
finite because of supersymmetric cancellations
in group theory factors:
$
\str \A_\mu = \str \one = 0
$.
Transverse parts of such
diagrams with four external legs are finite by power counting,
whilst the longitudinal parts are finite once gauge invariance properties
are taken into account \cite{sunn}. 

Finally, we need to show that at energies much lower than the cutoff,
the theory we are supposed to be regularising is recovered, namely
$SU(N)$ Yang-Mills. (In the ERG context the cutoff in question
is $\Lambda_0\to\infty$ where  
explicitly or implicitly, the partition function is defined.)
There is a case to answer because the massless
sector that remains, contains the second gauge field, $A^2$. In fact
this gauge 
field is unphysical because the supertrace in (\ref{eq:str})
gives it a wrong sign action, as can be seen from 
eq.~(\ref{eq:la}),\footnote{The supertrace is a necessity 
since it is this, 
not the trace, that leads to invariants when supergroups are used \cite{bars,sunn}.} 
leading to negative norms in its Fock space \cite{sunn}. Fortunately, the 
Appelquist-Carazzone theorem saves the day: since the $A^1$ and $A^2$ 
live in disjoint groups, the lowest dimension 
interaction between $A^1$ and $A^2$ is proportional to
$
\tr\left(F^1_{\mu\nu}\right)^2 \tr\left(F^2_{\mu\nu} \right)^2.
$
Since this is irrelevant, the $A^2$ sector decouples in the limit that
$\Lambda_0\to\infty$.

This completes the proof of finiteness to all orders of
perturbation theory, in four (or less) dimensions. 
In the limit $N=\infty$, the scheme can be shown to regularise in any 
dimension \cite{sunn}.

\section{Manifestly gauge invariant flow equation}
\subsection{Polchinski's equation}

We are ready to write a gauge invariant flow equation. To motivate it
consider Polchinski's version of Wilson's ERG \cite{polch}. We can cast it in
the form
\be
\label{eq:pol}
\ldl S = -{1\over \Lambda^2} \dphi{S}{} \cdot c' \cdot \dphi{\Sigma}{} +
{1\over \Lambda^2} \dphi{}{} \cdot c' \cdot \dphi{\Sigma}{}.
\ee
Here $\varphi$ is for example a single scalar field.
$\Sigma$ is the combination $S - 2 \hat{S}$, where 
$\hat{S}$ is the regularised kinetic
term $\hat{S}= {1\over 2} \demu \varphi \cdot c^{-1}\cdot \demu \varphi$.
In this form it is clear that the ERG leaves the partition function
invariant because the Boltzmann measure factor flows into a total 
functional derivative:
\be
\ldl \exp -S=  -{1\over \Lambda^2} \dphi{}{} \cdot c' \cdot \left(
\dphi{\Sigma}{} \exp -S \right).
\ee
At this stage we recognize that there is nothing particularly special
about the Polchinski / Wilson version. There are infinitely many other ERG 
flow equations with this property \cite{jose}, the continuum analogue of the
infinitely many possible blockings on the lattice. 
All we have to do is to choose a gauge invariant one by
making a gauge covariant replacement for 
$\Psi=c'\cdot\frac{\delta\Sigma}{\delta\varphi}$.

\subsection{\bSUN gauge invariant ERG}

Writing $\varphi\mapsto A_\mu$,
this can be done simply by replacing $\cdot c'\cdot$ with $\{c'\}$
and replacing $\hat{S}$ with a gauge invariant generalisation. Thus:
\be
\label{eq:ga}
\ldl S = -{1\over \Lambda^2} \frac{\delta S}{\delta A_\mu} \{ c' \}
\frac{\delta \Sigma_g}{\delta A_\mu} +
{1\over \Lambda^2} \frac{\delta }{\delta A_\mu} \{ c' \} \frac{\delta
\Sigma_g}{\delta A_\mu}  +\cdots,
\ee
where $\hat{S}= {1\over 4} F_{\mu \nu} \{ c^{-1}
\} F_{\mu \nu} +\cdots$. Recall that the coupling $g$
was scaled out, \cf eq.~(\ref{eq:1}). It must reappear somewhere in the
flow equation and some thought shows that the appropriate place is in
the combination $\Sigma_g = g^2 S - 2 \hat{S}$.
We have added the ellipsis in the recognition
that further regularisation will be needed over and above 
the gauge invariant higher derivatives.

\subsection{\bSUNN gauge invariant ERG}

We get the remaining regularisation by promoting $A$ to $\A$, adding
the super-scalar sector, and then shifting $\C$
to the fermionic symmetry breaking vacuum expectation value.
We want to ensure that under the ERG flow, this vacuum expectation
value flows with the effective cutoff, \ie as $<\!\C\!>=\Lambda\sigma_3$.
One can show that this follows at the classical level if we work instead
with a dimensionless superscalar, $\C\mapsto\C\Lambda$, 
so that the shift becomes $\C\rightarrow\C+\sigma_3$. It is technically
very convenient if the ERG equation allows for the classical two-point
vertices to be equal to those coming from $\hat{S}$ \cite{ymii}, as is true
(\ref{eq:pol}) and (\ref{eq:ga}) above. To keep this 
property in the spontaneously broken phase we need different kernels
for $B$ and $D$ which we can make by adding $\C$ commutator terms. 
Constructing the appropriate $\Psi$, we thus obtain in the symmetric
phase, a fully manifestly $SU(N|N)$ gauge invariant flow equation:
$
\ldl S =
- a_0[S,\Sigma_g]+a_1[\Sigma_g],
$
where
\ba
 a_0[S,\Sigma_g]
 = &&\frac{1}{2\Lam^2}\left(\!\frac{\delta S}{\delta {\cal
A}_{\mu}}\{c'\}\frac{\delta \Sigma_g}{\delta {\cal
A}_{\mu}}\!-\!\frac{1}{4}[{\cal C},\frac{\delta S}{\delta {\cal
A}_{\mu}}]\{{M}\}[{\cal C},\frac{\delta \Sigma_g}{\delta {\cal
A}_{\mu}}]\!\right) \nonumber\\
&&+\frac{1}{2{\Lam^4}}\left(\!
\frac{\delta S}{\delta {\cal C}}\{H\}\frac{\delta \Sigma_g}{\delta {\cal
C}}\!-\!\frac{1}{4}[{\cal C},\frac{\delta S}{\delta {\cal
C}}]\{{L}\}[{\cal C},\frac{\delta \Sigma_g}{\delta {\cal
C}}]\!\right) \nonumber\\
a_1[\Sigma_g] = &&\frac{1}{2 \Lam^2}\left(\frac{\delta }{\delta {\cal
A}_{\mu}}\{c'\}\frac{\delta \Sigma_g}{\delta {\cal
A}_{\mu}}-\frac{1}{4}[{\cal C},\frac{\delta }{\delta {\cal
A}_{\mu}}]\{M\}[{\cal C},\frac{\delta \Sigma_g}{\delta {\cal
A}_{\mu}}]\right)\nonumber\\
&&+\frac{1}{2 \Lam^4}\left(
\frac{\delta}{\delta {\cal
C}}\{H\}\frac{\delta \Sigma_g}{\delta \C}
-\frac{1}{4}[{\cal C},\frac{\delta }{\delta {\cal
C}}]\{L\}[{\cal C},\frac{\delta \Sigma_g}{\delta {\cal
C}}]\!\right).
\ea
In here, we can take $\hat{S} = \int\! {\rm d}^4x \left( {\cal L}_\A + 
{\cal L}_\C \right)$, although there is considerable flexibility over
the exact choice as there is with the covariantization, 
and recognising this, we were able to turn this to
our advantage \cite{sa,next}.
The kernels $M$, $H$ and $L$ are then determined in terms
of $c$, $\ct$ and other parameters in $\hat{S}$ (here $\lambda$)
by the requirement
that the classical solution $S$ can have the same two-point vertices 
as $\hat{S}$.
Although $g$ appears explicitly as a parameter in these flow equations, 
it is not yet defined as the running Yang-Mills coupling. As usual, this 
is done via a renormalisation condition: for the pure $A^1$ part we 
require
\be
S = \frac{1}{2 g^2(\Lam)} \,\tr \int\!\! d^4\! x\, (
F_{\mu\nu}^1 )^2  +{O}(D^3).
\ee

At first sight, it appears that we have specialized the kernels for the
gauge fields so that no longitudinal terms appear. In fact, any
longitudinal term $\sim\nabla_\mu\cdot 
{\delta S\over\delta\A_\mu}$ may be converted to $\C$ commutator terms,
$\C \cdot \frac{\delta S}{\delta \C}$,
\ie $L$ type terms, via $SU(N|N)$ gauge invariance. 

The supermatrix functional derivatives are most easily computed by
noting that they have a very simple effect on supertraces. Either
we have `supersowing',
$
\str A\frac{\delta}{\delta X} \str XB = \str AB,
$
or `supersplitting',
$
\str \frac{\delta}{\delta X} AXB=\str A \, \str B.
$
Drawing single supertraces as closed curves, and using Fig.~\ref{fig:2},
we get a useful diagrammatic interpretation 
which counts supertraces, analogous to the 't Hooft double-line
notation \cite{thooft}.
These relations follow from the completeness relations for the
supergenerators. Just as in the analogous formulae for $SU(N)$, 
generically there are $1/N$ corrections, but
they involve ordinary traces (or equivalently $\sigma_3$) 
which would violate $SU(N|N)$. In the case of this
$SU(N|N)$ gauge theory, they must all cancel out and they do \cite{sunn,next},
so the double line notation is exact -- even at finite $N$ \cite{next}.

Finally, shifting $\C$ to $\C+\sigma_3$, allows us to perform computations
in which not only unbroken $SU(N)\times SU(N)$ gauge invariance,
but also broken fermionic gauge
invariance, is manifestly preserved at every step. As well as providing
yet another beautiful balance in the formalism, one sees
very clearly how a massive vector field ($B$) as created by spontaneous
symmetry breaking, and its associated Goldstone mode ($D$), actually form 
a single unit, tied together by the underlying gauge invariance \cite{next}. 

\section{Summary and conclusions}

A manifestly gauge invariant ERG, together with the
necessary non-perturbative gauge invariant regularisation scheme,
has been proposed. No gauge fixing is required to define it, and as we
will see in the next talk no gauge fixing is needed to compute the 
solutions either \cite{alg,ymi,ymii}, thus avoiding the Gribov problem 
\cite{gribov}.
Although there has been no room to explain it, 
especially the gauge sector (\ref{eq:ga}), may be
reinterpreted in terms of Wilson loops, the natural order parameter 
for gauge theory. The ERG then has an
interpretation in the large $N$ limit as
quantum mechanics of a single Wilson loop, with close links to
the Migdal-Makeenko equation \cite{ymi,mig}.

As explained in the next lecture, we are in the midst of developing and
testing perturbatively, powerful methods of computation in gauge theory
based on the above ERG.
For the future, we intend to include matter in the fundamental 
representation, and turn our attention to non-perturbative 
approximations and QCD. It also seems a simple matter to incorporate
space-time supersymmetry, opening up intruiging possibilities for 
deeper investigations of Seiberg-Witten methods and the 
AdS/CFT correspondence \cite{saky,sw,adscft}.

\begin{ack}
The authors wish to thank the organisers of the fifth international
Conference ``RG 2002'' for providing such a stimulating environment. 
T.R.M. and S.A. acknowledge financial support from PPARC Rolling Grant
PPA/G/O/2000/00464.
\end{ack}

\end{document}